\pdfoutput=1

\documentclass[11pt,twoside,a4paper,cmspaper,final,collab]{cms-tdr}
\def\svnVersion{03f6a75-D}\def\svnDate{2019/10/02}\def\cmsCernNoTag{CERN-EP-2019-089}\def\cmsCernDate{\today}\def\cmsMessage{"Published in Physics Letters B as \href{http://dx.doi.org/10.1016/j.physletb.2019.134985}{\doi{10.1016/j.physletb.2019.134985}.}"}

\begin{document}\cmsNoteHeader{SMP-18-006}

\hyphenation{had-ron-i-za-tion}
\hyphenation{cal-or-i-me-ter}
\hyphenation{de-vices}
\RCS$HeadURL$
\RCS$Id$

\newlength\cmsFigWidth
\ifthenelse{\boolean{cms@external}}{\setlength\cmsFigWidth{0.85\textwidth}}{\setlength\cmsFigWidth{0.4\textwidth}}
\newlength\cmsFigWidthSplit
\ifthenelse{\boolean{cms@external}}{\setlength\cmsFigWidthSplit{0.49\textwidth}}{\setlength\cmsFigWidthSplit{0.49\textwidth}}
\ifthenelse{\boolean{cms@external}}{\providecommand{\cmsLeft}{upper\xspace}}{\providecommand{\cmsLeft}{left\xspace}}
\ifthenelse{\boolean{cms@external}}{\providecommand{\cmsRight}{lower\xspace}}{\providecommand{\cmsRight}{right\xspace}}
\ifthenelse{\boolean{cms@external}}{\providecommand{\NA}{\ensuremath{\cdots}\xspace}}{\providecommand{\NA}{\ensuremath{\text{---}}\xspace}}
\ifthenelse{\boolean{cms@external}}{\providecommand{\cmsTableResize[1]}{\relax{#1}}}{\providecommand{\cmsTableResize[1]}{\resizebox{\columnwidth}{!}{#1}}}
\ifthenelse{\boolean{cms@external}}{\providecommand{\cmsFigOneABC}{upper-left, upper-right, lower-left}}{\providecommand{\cmsFigOneABC}{left, middle-left, middle-right}}
\ifthenelse{\boolean{cms@external}}{\providecommand{\cmsFigOneD}{lower-right}}{\providecommand{\cmsFigOneD}{right}}
\newcommand{\x}{\ensuremath{\phantom{0}}}
\newcommand{\y}{\ensuremath{\phantom{.}}}
\newcommand{\PHpmpm}{\ensuremath{\PH^{\pm\pm}}}
\providecommand{\PV}{\ensuremath{\mathrm{V}}}
\newlength\cmsTabSkip\setlength{\cmsTabSkip}{1ex}

\cmsNoteHeader{SMP-18-006}
\title{Search for anomalous electroweak production of vector boson pairs in association with two jets in proton-proton collisions at 13\TeV}

\date{\today}

\abstract{
A search for anomalous electroweak production of $\PW\PW$, $\PW\PZ$, and $\PZ\PZ$ boson pairs in association with two jets in proton-proton collisions at $\sqrt{s}=13\TeV$ at the LHC is reported. The data sample corresponds to an integrated luminosity of $35.9\fbinv$ collected with the CMS detector. Events are selected by requiring two jets with large rapidity separation and invariant mass, one or two leptons (electrons or muons), and a $\PW$ or $\PZ$ boson decaying hadronically. No excess of events with respect to the standard model background predictions is observed and constraints on the structure of quartic vector boson interactions in the framework of dimension-8 effective field theory operators are reported. Stringent limits on parameters of the effective field theory operators are obtained. The observed $95\%$ confidence level limits for the S0, M0, and T0 operators are $-2.7<f_{\text{S0}}/ \Lambda^{4}<2.7$, $-1.0<f_{\text{M0}}/ \Lambda^{4}<1.0$, and $-0.17<f_{\text{T0}}/ \Lambda^{4}<0.16$, in units of \TeV{}$^{-4}$. Constraints are also reported on the product of the cross section and branching fraction for vector boson fusion production of charged Higgs bosons as a function of mass from 600 to 2000\GeV. The results are interpreted in the context of the Georgi--Machacek model.
}

\hypersetup{%
pdfauthor={CMS Collaboration},%
pdftitle={Search for anomalous electroweak production of vector boson pairs in association with two jets in proton-proton collisions at 13 TeV},%
pdfsubject={CMS},%
pdfkeywords={CMS, physics, VBS, charged Higgs}}

\maketitle
\section{Introduction}\label{sec:intro}

\begin{figure*}[htb]
\centering
\includegraphics[width=0.49\textwidth]{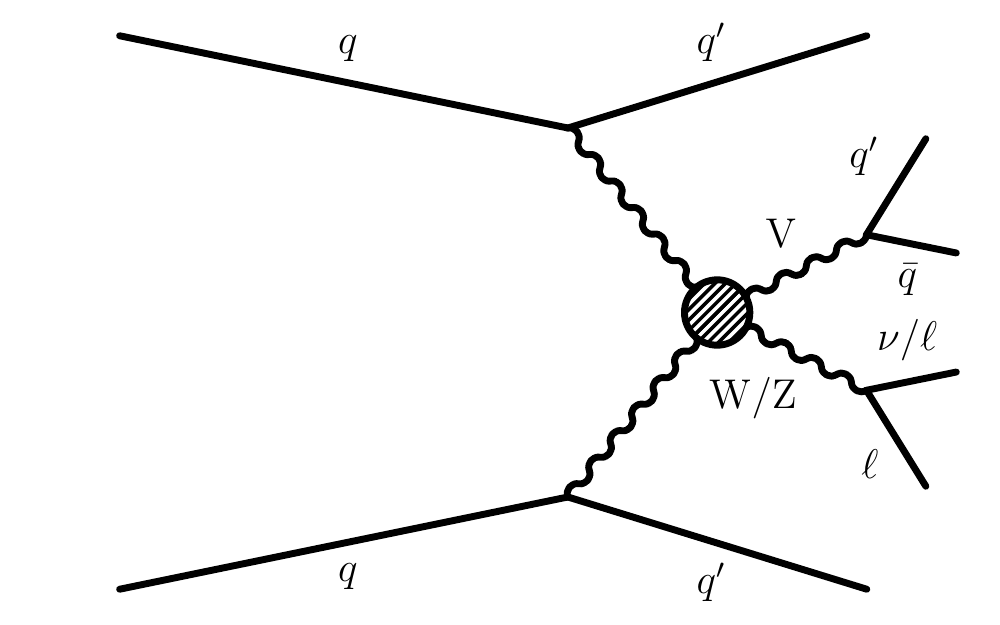}
\caption{The Feynman diagram of a VBS process contributing to the EW-induced production of events containing a hadronically decaying gauge boson ($\PV$), a $\PW^{\pm}$/$\PZ$ boson decaying to leptons, and two forward jets. New physics (represented by a black circle) in the EW sector can modify the quartic gauge couplings.}
\label{fig:feynman}
\end{figure*}

Measurements of vector boson scattering (VBS) processes probe the non-Abelian gauge structure of the electroweak (EW) interactions of the standard model (SM) of particle physics. The non-Abelian structure of the EW sector leads to self-interactions between gauge bosons via triple and quartic gauge couplings. At the CERN LHC interactions from VBS are characterized by the presence of two gauge bosons in association with two forward jets with large rapidity separation and large dijet invariant mass. The discovery of a Higgs boson~\cite{Aad:2012tfa,Chatrchyan:2012xdj,CMSPaperCombinationLong} established that $\PW$ and $\PZ$ gauge bosons acquire mass via the Higgs mechanism. Models of physics beyond the SM predict enhancements in VBS processes through modifications of the Higgs boson couplings to gauge bosons~\cite{PhysRevD.87.055017, PhysRevD.87.093005}. Figure~\ref{fig:feynman} shows a representative Feynman diagram involving quartic vertices. An excess of events with respect to the SM predictions could indicate the presence of anomalous quartic gauge couplings (aQGCs)~\cite{aqgc_operators}.

This paper presents a study of VBS in $\PW\PW$, $\PW\PZ$, and $\PZ\PZ$ channels using proton-proton ($\Pp\Pp$) collisions at $\sqrt{s}=13\TeV$. The data sample corresponds to an integrated luminosity of $35.9 \pm 0.9\fbinv$~\cite{LUM-17-001} collected with the CMS detector~\cite{Chatrchyan:2008zzk} at the LHC in 2016.

The first goal of this paper is to search for the presence of aQGCs in candidate events containing a (\romannumeral 1) hadronically decaying gauge boson ($\PV$) produced with large transverse momentum $\pt$, (\romannumeral 2) a $\PW$ or $\PZ$ boson decaying to one or two charged leptons (electrons or muons), and (\romannumeral 3) two forward jets. This final state has a higher branching fraction of the $\PV$ decay than previous aQGC searches at the LHC for VBS containing only leptonic boson decays~\cite{Sirunyan:2017ret, ATLAS_ssWW,CMS_ssWW,Aaboud:2016ffv,Sirunyan:2019ksz,Aad:2016ett,Sirunyan:2017fvv,Khachatryan:2016mud,Khachatryan:2017jub,Khachatryan:2016vif,Aaboud:2017pds}. A $\PW \PV$ final state where the $\PW$ boson decays to leptons receives contributions from the production of $\PW^{\pm}\PW^{\mp}$, $\PW^{\pm}\PW^{\pm}$, and $\PW^{\pm}\PZ$ boson pairs. Similarly, a $\PZ \PV$ final state where the $\PZ$ boson decays to leptons receives contributions from the production of $\PW^{\pm}\PZ$ and $\PZ\PZ$ boson pairs. The ATLAS and CMS Collaborations have reported limits on aQGCs using final states with a hadronically decaying $\PW/ \PZ$ boson in $\Pp\Pp$ collisions at center-of-mass energy $\sqrt{s}=8\TeV$~\cite{Aaboud:2016uuk,Aaboud:2016ftt,Aaboud:2017tcq,Chatrchyan:2014bza}.

A second goal of this paper is to search for charged Higgs bosons that are produced via vector boson fusion (VBF) and decay to $\PW$ and $\PZ$ bosons. Proposals exist for extended Higgs sectors with additional SU(2) isotriplet scalars that give rise to charged Higgs bosons with couplings to $\PW$ and $\PZ$ bosons at the tree-level~\cite{CE1,CE2}. Specifically, the Georgi--Machacek (GM) model~\cite{GEORGI1985463}, with both real and complex triplets, preserves a global symmetry SU$_\mathrm{L}$(2)$\times$SU$_\mathrm{R}$(2), which is broken by the Higgs vacuum expectation value to the diagonal subgroup SU$_{\mathrm{L}+\mathrm{R}}$(2). Thus, the tree-level ratio of the $\PW$ and $\PZ$ boson masses is protected against large radiative corrections. In this model, singly (doubly) charged Higgs bosons are produced via VBF that decay to $\PW$ and $\PZ$ bosons (same-sign $\PW$ boson pairs).

\begin{figure*}[htb]
\centering
\includegraphics[width=0.49\textwidth]{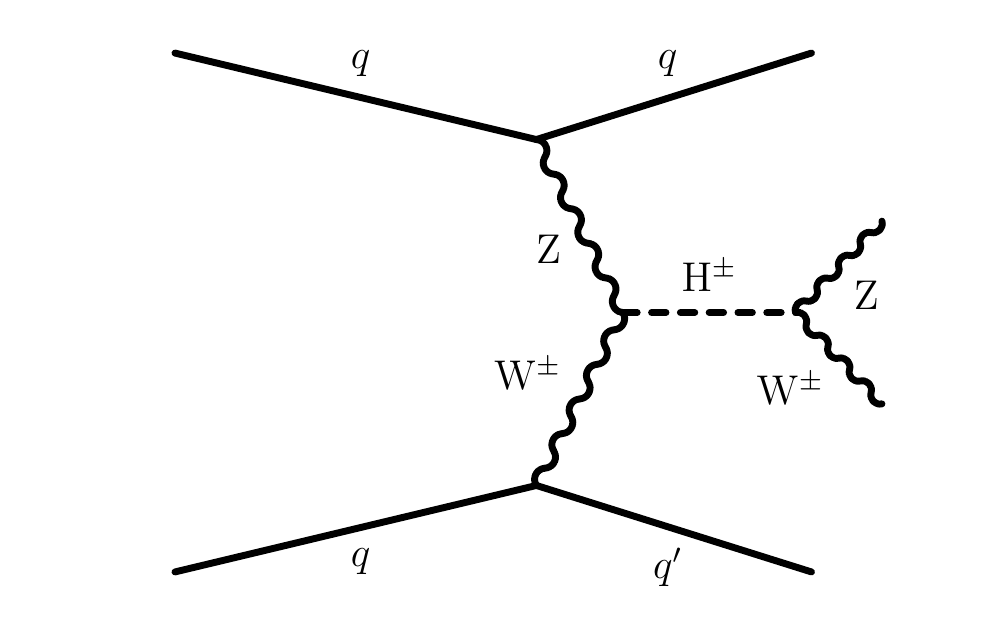}
\includegraphics[width=0.49\textwidth]{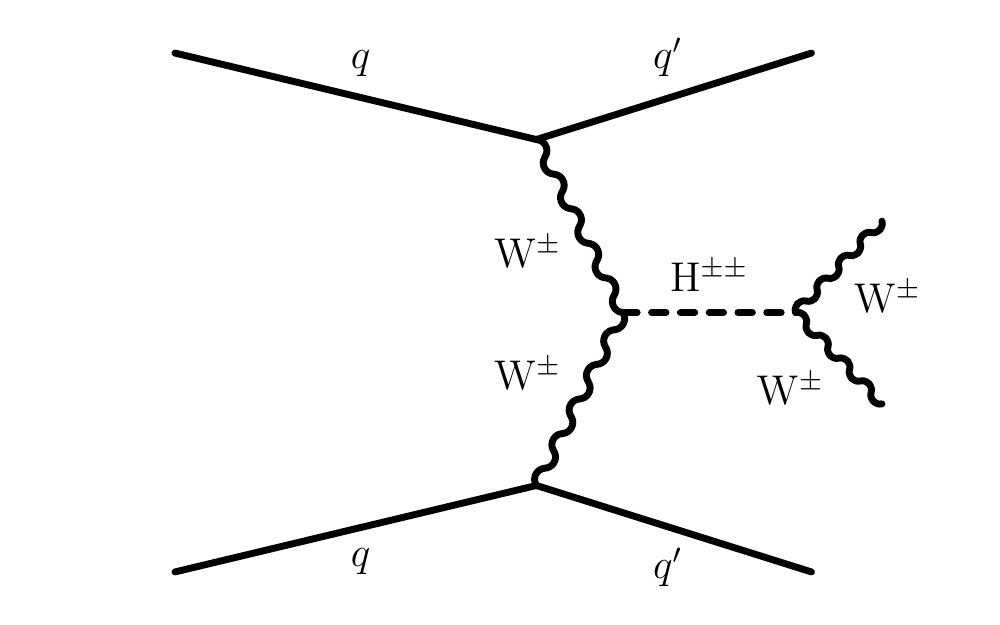}
\caption{Examples of Feynman diagrams showing the production of singly (left) and doubly (right) charged Higgs bosons via VBF.}
\label{fig:feynman2}
\end{figure*}

The charged Higgs bosons $\PHpm$ and $\PHpmpm$ in the GM model are degenerate in mass (denoted as $m(\PH_{5})$) at tree level and transform as a quintuplet under the SU$_{\mathrm{L}+\mathrm{R}}$(2) symmetry. The coupling depends on $m(\PH_{5})$ and the parameter $s_{\PH}$, where $s_{\PH}^2$ characterizes the fraction of the $\PW$ boson mass squared generated by the vacuum expectation value of the triplet fields. Figure~\ref{fig:feynman2} shows representative Feynman diagrams for the production and decay of the charged Higgs bosons. The CMS Collaboration at $13\TeV$~\cite{Sirunyan:2017ret,Sirunyan:2017sbn,Sirunyan:2019ksz} and the ATLAS Collaboration at $8\TeV$~\cite{PhysRevLett.114.231801} performed searches for charged Higgs bosons in these topologies and set constraints on the GM model.

\section{The CMS detector}\label{sec:cms}
The central feature of the CMS apparatus is a superconducting solenoid
of 6\unit{m} internal diameter, providing a magnetic field of
3.8\unit{T}. Within the solenoid volume are a silicon pixel and strip
tracker, a lead tungstate crystal electromagnetic calorimeter (ECAL),
and a brass and scintillator hadron calorimeter, each composed
of a barrel and two endcap sections. Forward calorimeters extend the pseudorapidity ($\eta$) coverage provided by the barrel and endcap detectors. Muons are detected in gas-ionization detectors embedded in the steel flux-return yoke outside the solenoid. A more detailed description of the CMS detector, together with a definition of the coordinate system used and the relevant kinematic variables, can be found in Ref.~\cite{Chatrchyan:2008zzk}.

The first level of the CMS trigger system, composed of custom hardware processors, uses information from the calorimeters and muon detectors to select events of interest in a fixed time interval of less than 4\mus. The second level, known as the high-level trigger, consists of a farm of processors running a version of the full event reconstruction software optimized for fast processing, and reduces the event rate to $\mathcal{O}$(1\unit{kHz}) before data storage~\cite{Khachatryan:2016bia}.

\section{Signal and background simulation}\label{sec:event_reco}
The SM EW, aQGC, and charged Higgs boson processes with two final-state quarks are simulated using the Monte Carlo (MC) generator \MADGRAPH{}5\_a\MCATNLO~2.3.3~\cite{Alwall:2014} at leading order (LO) with four EW and zero quantum chromodynamic (QCD) vertices. The signatures of $\PW^{\pm}\PW^{\pm}$, $\PW^{\pm}\PW^{\mp}$, $\PW^{\pm}\PZ$, and $\PZ\PZ$ processes are produced separately and include diagrams with quartic vertices. The simulation of the aQGC processes employs matrix element reweighting to obtain a finely spaced grid of parameters for each of the anomalous couplings probed by the analysis.

The production of two gauge bosons with two final state quarks or gluons and at least one QCD vertex at tree level, which is referred to as QCD $\PV \PV$ production, is considered background. The \MADGRAPH{}5\_a\MCATNLO~2.3.3 generator at LO is used to simulate this process. The interference between the EW and QCD diagrams is evaluated using dedicated samples produced with the \textsc{Phantom}~1.2.8~\cite{Ballestrero:2007xq} generator. The effect of the interference contributes at the level of 1\% in the signal region and is, therefore, neglected.

The $\PW$+jets and Drell--Yan processes, with up to four outgoing partons at Born level, are simulated at QCD LO accuracy using \MADGRAPH{}5\_a\MCATNLO. The $\ttbar$, $\ttbar \PW$, $\ttbar \PZ$, and single top quark processes are generated at next-to-leading order (NLO) accuracy using \textsc{POWHEG}~2.0~\cite{Alioli:2008gx,Nason:2004rx,Frixione:2007vw,powheg:2010}. The simulated samples of background processes are normalized to the best prediction available, NLO or higher~\cite{Gavin:2012sy,Li:2012wna,Gavin:2010az,bib:twchan,Czakon:2011xx}.

The \PYTHIA~8.212~\cite{Sjostrand:2015} package with the tune CUETP8M1~\cite{Skands:2014pea,Khachatryan:2015pea} is used for parton showering, hadronization, and the underlying event simulation. The NNPDF~3.0~\cite{nnpdf} set is used as the default set of parton distribution functions (PDFs). The PDFs are calculated at the same order as the corresponding hard process.

The detector response is simulated using a detailed description of the CMS detector based on the \GEANTfour package~\cite{Agostinelli:2002hh}, and event reconstruction is performed with the same algorithms used for data. Additional $\Pp\Pp$ interactions (pileup) occurring in the same beam crossing as the event of interest are included in the simulation. These events are weighted so that the pileup distribution matches that observed in data, which has an average of approximately 23 interactions per beam crossing assuming $69\unit{mb}$ for the inelastic $\Pp\Pp$ cross section~\cite{Sirunyan:2018nqx}.

\section{Event reconstruction and selection}\label{sec:event_sel}

The particle-flow algorithm~\cite{Sirunyan:2017ulk} reconstructs and identifies each individual particle in an event, with an optimized combination of all subdetector information. The individual particles are identified as charged and neutral hadrons, leptons, and photons. The missing transverse momentum, $\ptvecmiss$, is defined as the magnitude of the negative vector $\pt$ sum of all reconstructed particles in the event. Its magnitude is denoted by $\ptmiss$.

Jets are reconstructed using the anti-\kt clustering algorithm~\cite{antikt}  with a distance parameter of 0.4, as implemented in the \FASTJET package~\cite{Cacciari:fastjet1,Cacciari:fastjet2}. Jet momentum is determined as the sum of all particle momenta in the jet. Corrections are applied to the jet energy as a function of jet $\eta$ and $\pt$ to account for detector response nonlinearities, contribution from pileup, and residual differences between the jet energy scale in data and simulation~\cite{1748-0221-6-11-P11002, Khachatryan:2016kdb}. Additional selection requirements remove spurious jets originating from isolated noise patterns in certain regions of the hadron calorimeter~\cite{Sirunyan:2019kia}. These corrections are also propagated to the $\ptmiss$ calculation. The {\cPqb} quark jet identification criteria are based on a multivariate technique to combine the information from displaced tracks with the information from secondary vertices associated with the jet and on the possible presence of a soft muon in the event from the semileptonic decay of the {\cPqb} quark~\cite{btag}.

High-energy $\PV$ boson candidates, referred to as $\PV$ jets, are reconstructed using the anti-\kt clustering algorithm~\cite{antikt} with a distance parameter of 0.8~\cite{Khachatryan:2014vla}. The \textsc{puppi} algorithm~\cite{Bertolini:2014bba} is used to mitigate the effect of pileup by assigning a weight to each particle prior to jet clustering based on the likelihood of the particle originating from pileup. The mass of the $\PV$ jet ($m_{\PV}$) is computed after employing the modified mass-drop tagger algorithm~\cite{Dasgupta:2013ihk,Larkoski:2014wba} to remove soft, wide-angle radiation from the jets. The $N$-subjettiness variable $\tau_N$~\cite{Thaler:2010tr} quantifies how well the jet can be divided into $N$ subjets. The observable $\tau_2 / \tau_1$ is employed to discriminate 2-prong objects arising from hadronic decays of $\PW$ or $\PZ$ bosons from those from light quarks or gluons.

The reconstructed vertex with the largest value of summed physics-object $\pt^2$ is the primary $\Pp\Pp$ interaction vertex. The physics objects are the jets, clustered using the jet-finding algorithm~\cite{antikt,Cacciari:fastjet1} with the tracks assigned to the vertex as inputs, and the associated missing transverse momentum, the negative vector sum of the \pt of those jets.

Muons are reconstructed by associating a track reconstructed in the inner silicon detectors with a track in the muon system. Selected muon candidates are required to satisfy a set of quality requirements based on the number of spatial measurements in the silicon tracker and the muon system, as well as the fit quality of the combined muon track~\cite{Chatrchyan:2012xi,Sirunyan:2018fpa}.

Electrons are reconstructed by associating a track reconstructed in the inner silicon detectors with a cluster of energy in ECAL~\cite{Khachatryan:2015hwa}. The selected electron candidates cannot originate from photon conversions in the inner silicon tracker material and must satisfy a set of quality requirements based on the shower shape of the energy deposit in the ECAL. Electron candidates in the transition region between the ECAL barrel and endcap, $1.44 < \abs{\eta} < 1.57$, are not considered because this transition region leads to lower quality reconstructed clusters because of a gap between the barrel and endcap calorimeters, which is filled with services and cables.

The lepton candidate tracks must be consistent with the primary vertex of the event~\cite{TRK-11-001} to suppress electron candidates from photon conversions and lepton candidates originating from decays of heavy quarks. The lepton candidates must be isolated from other particles in the event. The relative isolation for the lepton candidates with transverse momentum $\pt^{\ell}$ is defined as
\begin{linenomath}
\begin{equation}
R_\text{iso} = \bigg[\sum_{\substack{\text{charged} \\ \text{hadrons}}} \!\! \pt \, + \,
\max\big(0, \sum_{\substack{\text{neutral} \\ \text{hadrons}}} \!\! \pt
\, + \, \sum_{\text{photons}} \!\! \pt \, - \, \pt^\mathrm{PU}
\big)\bigg] \bigg/ \pt^{\ell},
\label{eq:iso}
\end{equation}
\end{linenomath}
where the sums run over the charged and neutral hadrons and photons in a
cone defined by
$\Delta R \equiv \sqrt{\smash[b]{\left(\Delta\eta\right)^2 + \left(\Delta\phi\right)^2}} = 0.4$ (0.3)
around the muon (electron) trajectory, and $\pt^\mathrm{PU}$ denotes the contribution of neutral particles from pileup~\cite{Chatrchyan:2012xi,Khachatryan:2015hwa}. Only charged hadrons originating from the primary vertex are included in the first sum.

Muon (electron) candidates with $\Delta R < 0.15$ (0.06) are considered isolated. The lepton reconstruction and selection efficiencies are measured using ``tag-and-probe'' techniques with Drell--Yan events that provide an unbiased sample with high purity~\cite{CMS:2011aa}. The muon (electron) candidates have an average selection efficiency of 95 (70)\%.

The event selection identifies events with one or two leptons and a high-energy $\PV$ boson produced with VBS topology. The events are triggered by the presence of at least one muon with $\pt>24\GeV$ and $\abs{\eta}<2.4$, or at least one electron with transverse energy $\ET>27\GeV$ and $\abs{\eta}<2.5$. These triggered muons and electrons satisfy less restrictive isolation and quality requirements than the offline selection criteria.

In the offline analysis events with at least one isolated lepton with $\pt>50\GeV$ are accepted as candidates. The $\PW \PV \rightarrow \ell \nu \PV$ decays are characterized by a significant amount of $\ptmiss$ associated with the undetected neutrino. The Drell--Yan and QCD multijet background processes are reduced by requiring $\ptmiss>50$ (80)\GeV in the muon (electron) final state. Candidate events with a second opposite-charged and same flavor isolated lepton with $\pt>30\GeV$ select the $\PZ \PV \rightarrow \ell \ell \PV$ decays. The candidate $\PZ$ boson invariant mass must be within 15\GeV of the nominal $\PZ$ boson mass~\cite{Tanabashi:2018oca}. The presence of additional muons or electron, with $\pt>20\GeV$ and $\abs{\eta}<2.4\,(2.5)$ for muons (electrons), satisfying less restrictive selection requirements than the signal lepton candidate selection and with average selection efficiencies above 95\%~\cite{Sirunyan:2018fpa,Khachatryan:2015hwa}, is used as a condition to further reduce events from the top quark and triboson background processes. Events with no $\PZ$ boson candidate selected and with two or more leptons are rejected. Events with a selected $\PZ$ boson candidate and with three or more leptons are also rejected.

Events are required to have at least one $\PV$ jet with $\pt>200\GeV$, $\abs{\eta}<2.4$, $\tau_2/\tau_1 < 0.55$, and $65 < m_{\PV} < 105\GeV$. The $\PV$ jets that are within $\Delta R < 1.0$ of one of the identified leptons are excluded. The efficiency of the $N$-subjettiness and mass requirements for the signal events is about 70\%, while the probability of misidentifying a quark or a gluon jet as a $\PV$ jet is 5\%. The $\PV$ jet mass resolution is about 15\%. In the case of multiple $\PV$ jet candidates, the one with mass closest to the nominal $\PW$ boson mass~\cite{Tanabashi:2018oca} is selected.

Events are required to contain at least two jets with $\pt>30\GeV$ and $\abs{\eta}<5.0$, and $\Delta R(\mathrm{j},\PV)>0.8$. In the case of more than two jet candidates, the pair with the largest dijet mass is selected. The VBS topology is targeted by requiring a large dijet mass $m_{\mathrm{jj}}>800\GeV$ and a large pseudorapidity separation $\abs{\Delta \eta_{\mathrm{jj}}}>4.0$. Events having one or more identified {\cPqb} quark jets with $\pt>30\GeV$ and $\abs{\eta}<2.4$ are rejected, decreasing the number of top quark background events.

The longitudinal component of the neutrino momentum in $\PW \PV \rightarrow \ell \nu \PV$ events is estimated by constraining the mass of the charged lepton and neutrino system to be the nominal $\PW$ boson mass~\cite{Tanabashi:2018oca}. This is similar to the approach used in a previous CMS search~\cite{Sirunyan:2018iff}. The resulting quadratic equation is solved using $\ptvecmiss$ as an estimate of the neutrino transverse momentum. The solution with the closest match to the longitudinal component of the charged lepton momentum is selected. Only the real part is considered if no real solution is found. The momentum of the $\PW$ boson is then uniquely determined.

Additional selection criteria are employed to enhance the sensitivity to aQGCs in the $\PW \PV$ channel. The $\PW$ and $\PV$ bosons in the VBS and VBF topologies are mostly produced in the central rapidity region with respect to the two selected jets. Candidate events are required to have  $z_{\PV}^{*} < 0.3$ and $z_{\PW}^{*} < 0.3$, where $z_{x}^{*} = \abs{\eta_{x}-(\eta_{\mathrm{j1}}+\eta_{\mathrm{j2}})/2}/\abs{\Delta \eta_{\mathrm{jj}}}$ is the Zeppenfeld variable~\cite{Rainwater:1996ud}, $\eta_{x}$ is the pseudorapidity of a gauge boson, and $\eta_{\mathrm{j1}}$ and $\eta_{\mathrm{j2}}$ are the pseudorapidities of the two selected jets. In addition, events are required to have $\vartheta>1.0$, where $\vartheta = \mathrm{min}(\mathrm{min}(\eta_{\PW},\eta_{\PV})-\mathrm{min}(\eta_{\mathrm{j1}},\eta_{\mathrm{j2}}),\mathrm{max}(\eta_{\mathrm{j1}},\eta_{\mathrm{j2}})-\mathrm{max}(\eta_{\PW},\eta_{\PV}))$ is the boson centrality. The extraction of the signal yields is performed with a fit to the mass distribution of the $\PW \PV$ or $\PZ \PV$ system to statistically subtract the SM background contributions.

\section{Background estimation}\label{sec:bkg_estimation}
\begin{figure*}[htbp]
	 \centering
	 \begin{tabular}{cc}
	 \includegraphics[width=0.49\textwidth]{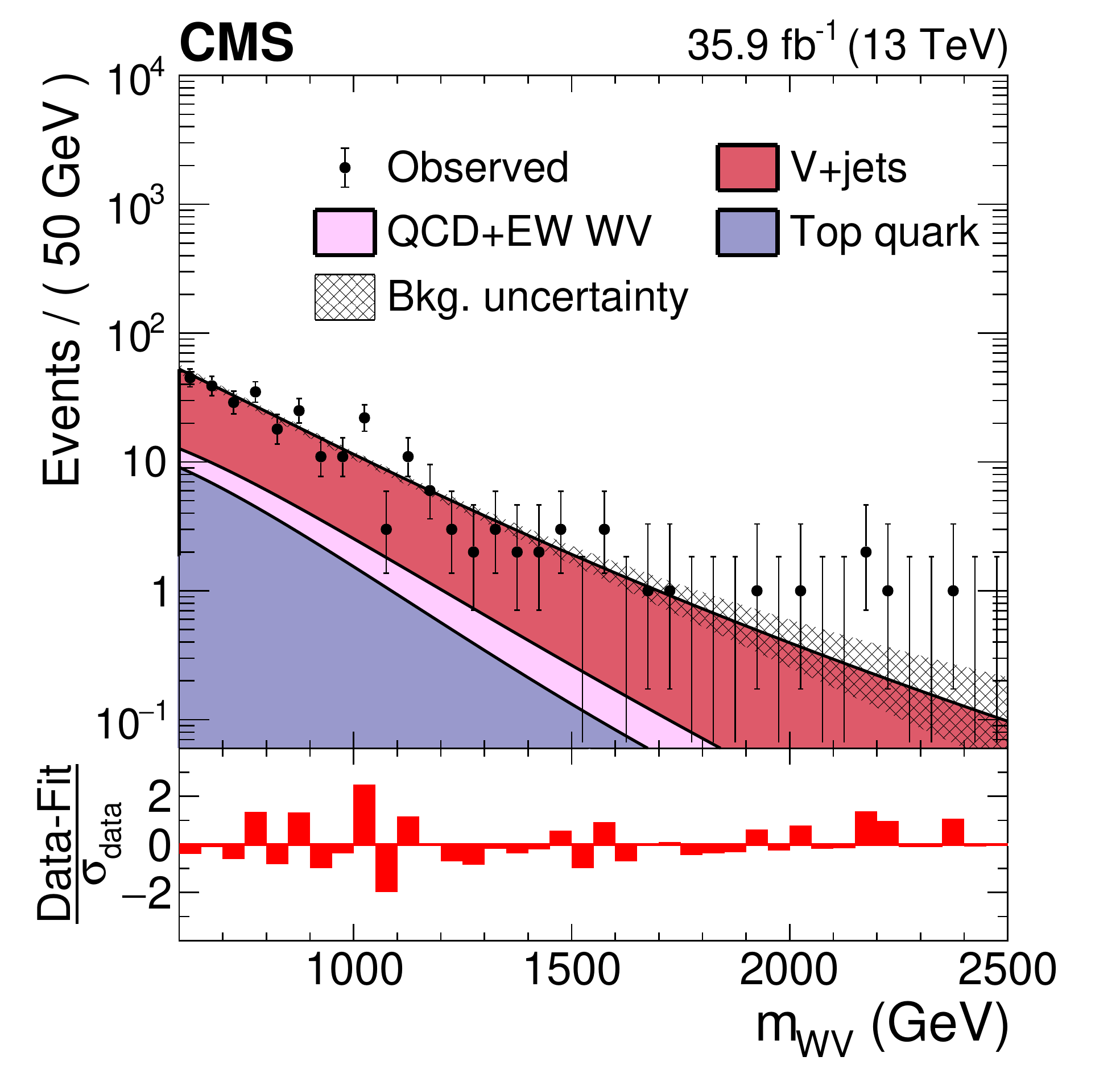}
	 \includegraphics[width=0.49\textwidth]{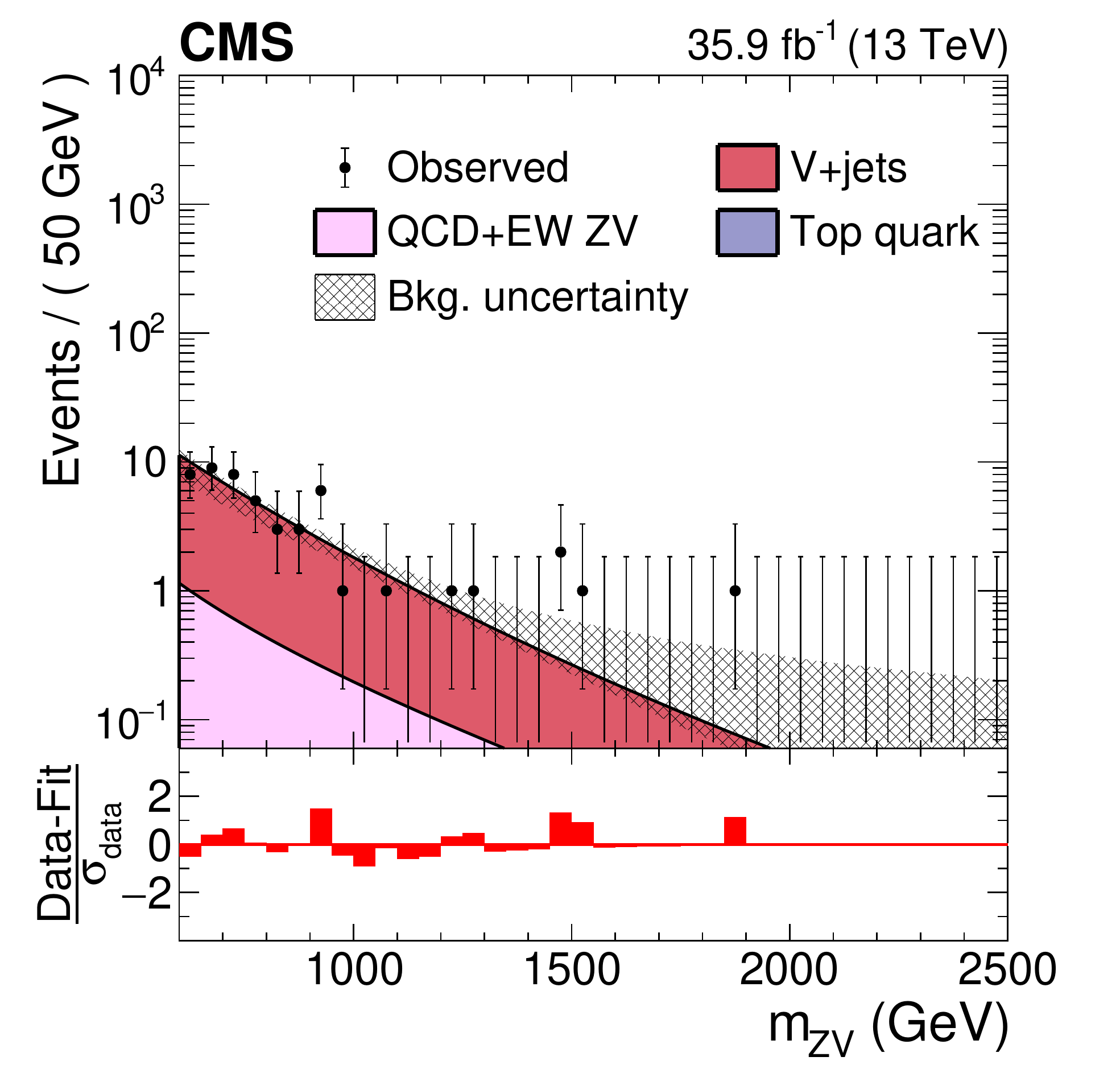}
	 \end{tabular}
	 \caption{Comparison between the fit results for the $\PV$+jets background processes and the data distributions of the $m_{\PW \PV}$ (left) and $m_{\PZ \PV}$ (right), respectively, in the sideband region with $40 < m_{\PV} < 65\GeV$ or $105 < m_{\PV} < 150\GeV$. The fit uncertainty is shown as a shaded band.}
	 \label{fig:DataMCForMWW}
\end{figure*}

The estimation of the shape and yield of the major background $\PW$($\PZ$)+jets in the $\PW \PV$ ($\PZ \PV$) channel is based on the observed data using the sideband of the signal region defined by the mass of the $\PV$ jet. The background estimation closely follows the methods used in Refs.~\cite{Gunion:1986cc,Khachatryan:2014gha,Sirunyan:2016cao}. An estimate of the $\PW$($\PZ$)+jets background is obtained by performing a maximum likelihood fit to the $m_{\PW \PV}$ ($m_{\PZ \PV}$) distribution in data for the events in the $\PW$($\PZ$)+jets enriched control region by selecting events with $40 < m_{\PV} < 65\GeV$ or $105 < m_{\PV} < 150\GeV$ and satisfying the rest of the signal selection criteria described in the last section. The background processes are modeled by fitting the $m_{\PW \PV}$ and $m_{\PZ \PV}$ distributions in the respective sideband regions with the parametric function $f(m)=\exp\left[-m/(c_{0}+c_1 m)\right]$. Figure~\ref{fig:DataMCForMWW} shows the $m_{\PW \PV}$ and $m_{\PZ \PV}$ data distributions and the corresponding fit results in this sideband region. The other background processes are also modeled by the parametric function in the fit with the shape and normalization fixed to the prediction from simulation. The SM EW $\PV \PV$ contribution is included in the fit. The contribution of the SM EW $\PV \PV$ process is expected to be small in the sideband region, even with enhancements of the cross section due to aQGCs, with a predicted yield of approximately $1\%$ of the selected events.

Transfer factors obtained from $\PW$($\PZ$)+jets simulation are used to extrapolate from the sideband to the signal region. The transfer factors are obtained from the ratio of the $\PW$($\PZ$)+jets yields in the signal and sideband regions as a function of the $m_{\PW \PV}$ ($m_{\PZ \PV}$). The statistical uncertainty in the transfer factor values due to the limited number of simulated events is also considered in the analysis and affects the normalization and shape of the $\PW$($\PZ$)+jets. The uncertainties in the fit parameters $c_{0}$ and $c_{1}$ are treated as nuisance parameters in the likelihood fit. The $\PW$($\PZ$)+jets estimation is also performed with an alternative function ($f(m)=\exp\left[-m/c_{0}\right]$) and the difference from the nominal prediction is taken as a systematic uncertainty.

The $m_{\PW \PV}$ ($m_{\PZ \PV}$) shapes of the $\ttbar$, $\ttbar \PW$, $\ttbar \PZ$, and single top quark background contributions in the signal region are predicted by the simulation after applying corrections to account for small differences between data and simulation~\cite{Sirunyan:2018fpa,Khachatryan:2015hwa,btag}. The event yields of these background processes are checked in a top quark enriched control sample by requiring a bottom quark jet in the final state. The QCD $\PV \PV$ background contribution is also evaluated from simulation.

\section{Systematic uncertainties}\label{sec:systematics}
A number of sources of systematic uncertainty can affect the rates and shapes of the $m_{\PW \PV}$ ($m_{\PZ \PV}$) distributions for the signal and background processes. Theoretical uncertainties are evaluated using the seven-point scale variation detailed in Ref.~\cite{Ballestrero:2018anz}, where the renormalization and factorization scales are varied independently up and down by a factor of two from their nominal value in each event (removing combinations where both variations differ by a factor of four). The largest variation from the nominal prediction is taken as a systematic uncertainty. The effect on the signal yields of the aQGC and charged Higgs bosons is up to 20\%, depending on the kinematic region. The effect on the expected yields of the SM EW $\PV \PV$ and QCD $\PV \PV$ processes reaches to 22 and 38\% for larger $m_{\PV\PV}$ values, respectively.

The PDF uncertainties are evaluated according to the procedure described in Ref.~\cite{PDFLHC} using the NNPDF~3.0~\cite{nnpdf} set. The uncertainty in the PDF results is up to 17\% variation for the signal, SM EW, and QCD $\PV \PV$ normalizations. The full NLO QCD and EW corrections for the SM EW and aQGC signal processes are not available and are not considered here. The NLO EW corrections are known only for the same-sign dilepton and $\PW\PZ\rightarrow \ell\ell'\ell'$ final states and reduce the cross section by approximately 15\%~\cite{Biedermann:2016yds,Biedermann:2017bss,Denner:2019tmn}. The uncertainty due to missing higher-order EW corrections in the GM model is evaluated to be 7\%~\cite{Zaro:2002500}.

The jet energy scale and resolution uncertainties affect the yields and shapes of the signal and background processes from simulation.  The effect on the expected yields reaches to above 10\% for larger $m_{\PV\PV}$ values. The uncertainties in the $\PV$ jet selection efficiency and $m_{\PV}$ scale and resolution give rise to a systematic uncertainty of 8\% in the predicted yields of the simulated processes. The lepton trigger, reconstruction, and selection efficiency uncertainties are 2.2 and 2.8\% for the $\PW\PV$ and $\PZ\PV$ channels, respectively. The {\cPqb} quark identification efficiency uncertainty results in 3\% systematic uncertainty in the top quark background normalization. An additional 5\% uncertainty is included for the top quark background normalization based on the level of agreement in yields between data and prediction in the {\cPqb} quark jet enriched control region. The uncertainty in the pileup reweighting uncertainty in the $\PV$ jet selection is evaluated by varying the effective inelastic cross section by 5\%~\cite{Sirunyan:2018nqx}. The statistical uncertainties due to the finite size of simulated samples are also included~\cite{BARLOW1993219}.

The $\PW$($\PZ$)+jets background normalization uncertainty is 7 (16)\%, dominated by the statistical uncertainty arising from the fit to the $m_{\PV\PV}$ distribution in the sideband region. The uncertainties in the fit parameters in the sideband region and the statistical uncertainty in the transfer factor values (described in Section~\ref{sec:bkg_estimation}) affect the shape of the $\PW$($\PZ$)+jets background distribution. Uncertainties affecting the $\PW$($\PZ$)+jets shapes are important for large $m_{\PV\PV}$ values reaching up to 200\%. The uncertainty of 2.5\% in the integrated luminosity determination~\cite{LUM-17-001} is included for all processes evaluated from simulation. This uncertainty does not affect the background processes estimated from data. A summary of the relative systematic uncertainties in the estimated signal and background yields is shown in Table~\ref{tab:HwzSystematics}.

\begin{table*}[htbp]
\topcaption{Relative systematic uncertainties in the estimated signal and background yields in units of percent. The range of the uncertainty variation as a function of $m_{\PV\PV}$ is shown for the systematic uncertainty sources affecting also the shape of the $m_{\PV\PV}$ distribution. The values in parenthesis show the systematic uncertainties in the $\PZ\PV$ channel where the uncertainties differ compared to the $\PW\PV$ channel.\label{tab:HwzSystematics}}
  \centering
\cmsTableResize{\begin{tabular}{lcccccc}
\hline
 Source & Shape & Signal (\%) & $\PV$+jets (\%)  & SM EW (\%) & QCD $\PV\PV$ (\%) & Top quark (\%)   \\

        \hline
Renorm./fact. scales      & $\checkmark$ & 11--22 &  \NA  &  11--22  &  32--38  & \NA 	 \\
PDF                       & $\checkmark$ & 7--17 & \NA &  4--17  & 5--9  & \NA	 \\
Jet momentum scale        & $\checkmark$ & 2--13  & \NA &  1--17  &  1--20  & 5--20	 \\
$\PV$ jet selection       &  & 8.0 & \NA &  8.0 &  8.0  & \NA 	 \\
GM model EW               &  & 7.0 & \NA &  \NA &  \NA  & \NA 	 \\
Bkg. normalization        & & \NA & 7 (16) &  \NA  &  \NA  &  5.0 \\
$\PV$+jets shape          & $\checkmark$  & \NA & 5--200 & \NA & \NA & \NA \\
Integrated luminosity     & & 2.5 &  \NA  &  2.5  &   2.5  & \NA \\
Lepton efficiency         & & 2.2 (2.8)  &  \NA  & 2.2 (2.8)  &  2.2 (2.8)  & \NA  \\
Lepton momentum scale     & $\checkmark$  & 0.5--3.5  & \NA &  0.5--3.5  &  1.5--7.5  & 1.0--5.0	 \\
{\cPqb} quark jet efficiency    & & 2.0 & \NA &  2.0 &   2.0  & 3.0 	 \\
Jet/$\ptmiss$ resolution  & & 4.0 & \NA &  3.0  & 2.0  & \NA \\
Pileup modeling           & & 4.0 & \NA & 4.0 & 4.0 & \NA \\
Limited MC event count    & $\checkmark$  & 1--2 & \NA & 6--20 (12--39) & 7--49 (17--57) & 5--50 (3--70) \\
\hline
      \end{tabular}}
\end{table*}

\begin{table*}[htbp]
\topcaption{Expected yields from various background processes in $\PW \PV$ and $\PZ \PV$ final states. The combination of the statistical and systematic uncertainties are shown. The predicted yields are shown with their best-fit normalizations from the background-only fit. The aQGC signal yields are shown for two aQGC scenarios with $f_{T2}/ \Lambda^{4} = -0.5\TeV^{-4}$ and $f_{T2}/ \Lambda^{4} = -2.5\TeV^{-4}$ for the  $\PW \PV$ and $\PZ \PV$ channels, respectively. The charged Higgs boson signal yields are also shown for values of $s_{\PH}=0.5$ and $m_{\PH_{5}}=500\GeV$ in the GM model. The statistical uncertainties are shown for the expected signal yields.
\label{tab:sel_yields15}}
\centering
\newcolumntype{x}{D{,}{\,\pm\,}{3.3}}
\begin{tabular}{lx{c}@{\hspace*{5pt}}x}
\hline
Final state   	      &   \multicolumn{1}{c}{$\PW \PV$} &&  \multicolumn{1}{c}{$\PZ \PV$}	       	  \\
\hline
Data                  &   \multicolumn{1}{c}{347}   && \multicolumn{1}{c}{47}               \\[\cmsTabSkip]
$\PV$+jets    	      &   196 ,  14  &&    42.6 ,  6.1    \\
Top quark             &   113 ,  15  &&    0.14 ,  0.04    \\
QCD $\PV \PV$  	      &   27 ,  8  &&    5.5 ,  1.9      \\
SM EW $\PV \PV$       &   16 ,  2  &&    2.0 ,  0.4      \\[\cmsTabSkip]
Total bkg.            &   352 ,  19   &&    50.3 ,  5.8   \\[\cmsTabSkip]
$f_{\text{T2}}/ \Lambda^{4} = -0.5,-2.5\TeV^{-4}$    &   19 , 1 && 6.7 , 0.5\\
$m_{\PH_{5}}=500\GeV$, $s_{\PH}=0.5$      &   38 , 1 && 4.1 , 0.1\\
\hline
\end{tabular}
\end{table*}

\section{Results}\label{sec:results}
No excess of events with respect to the SM background predictions is observed. The events in the signal region are used to constrain aQGCs in the effective field theory framework~\cite{Degrande:2012wf}. Nine independent charge conjugate and parity conserving dimension-8 effective operators are considered~\cite{aqgc_operators}. The S0 and S1 operators are constructed from the covariant derivative of the Higgs doublet. The T0, T1, and T2 operators are constructed from the SU$_\mathrm{L}$(2) gauge fields. The mixed operators M0, M1, M6, and M7 involve the  SU$_\mathrm{L}$(2) gauge fields and the Higgs doublet.

\begin{figure*}[htb]
\centering
\includegraphics[width=0.49\textwidth]{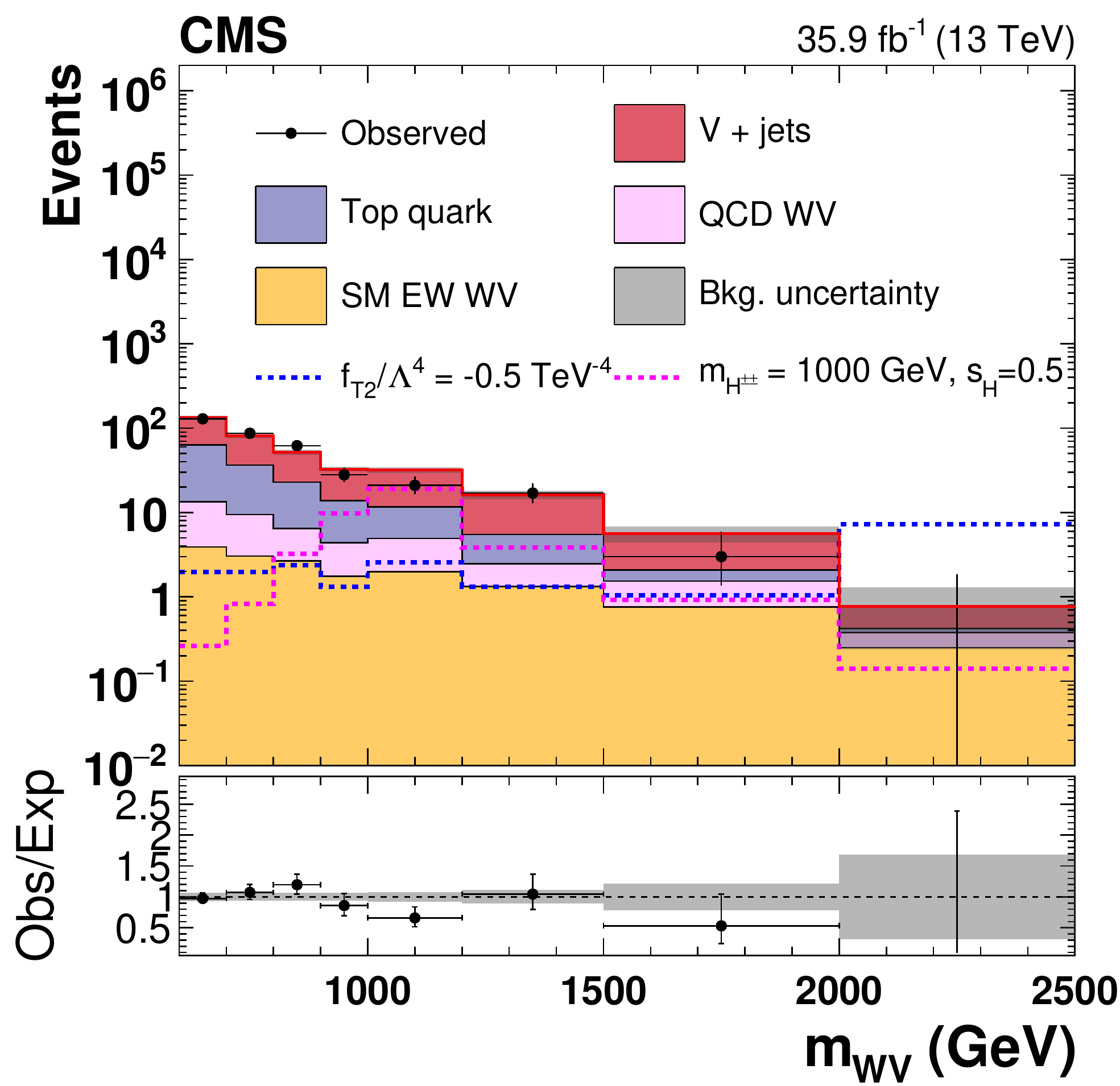}
\includegraphics[width=0.49\textwidth]{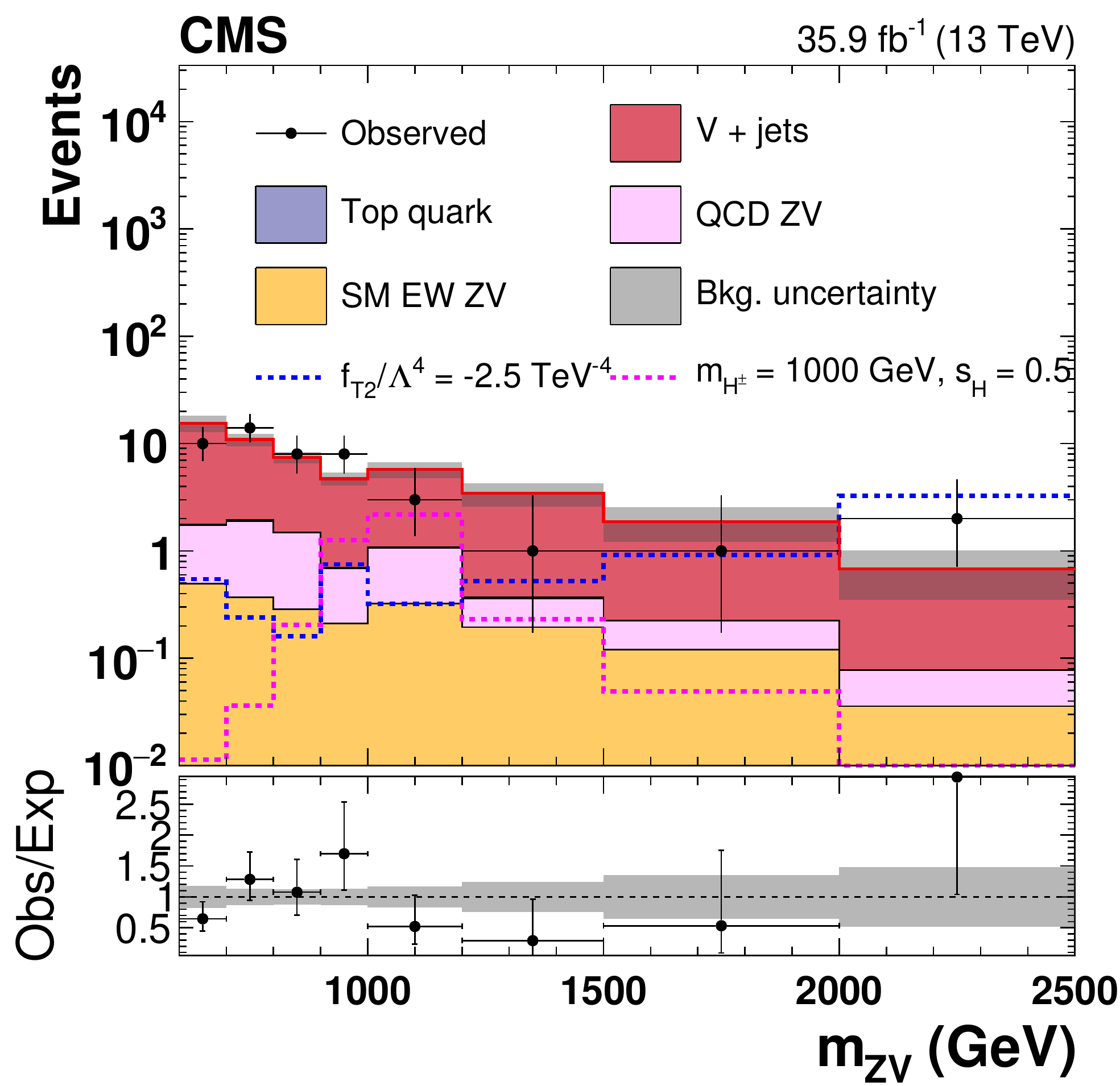}
\caption{Distributions of $m_{\PW \PV}$ (left) and $m_{\PZ \PV}$ (right) in the signal region. The gray bands include uncertainties from the predicted yields. The histograms for other backgrounds include the contributions from QCD $\PV \PV$, top quark, $\PW$+jets, and Drell--Yan processes. The predicted yields are shown with their best-fit normalizations from the background-only fit. The dashed lines show the signal predictions for two aQGC parameters, and charged Higgs bosons in the GM model. The overflow is included in the last bin. The bottom panel in each figure shows the ratio of the number of events observed in data to that of the total background prediction.}
\label{fig:signal2}
\end{figure*}

Statistical analysis of the event yields is performed with a fit to the mass distribution of the $\PW \PV$ or $\PZ \PV$ system in the signal region. The systematic uncertainties are treated as nuisance parameters in the fit and profiled. The SM EW production is treated as a background in the statistical analysis. The mass distributions are binned as follows: $m_{\PV \PV}$ = [600, 700, 800, 900, 1000, 1200, 1500, 2000, $\infty$]$\GeV$. The bin boundaries are chosen based on the limited number of simulated events for the background processes evaluated from simulation. The distributions of $m_{\PW \PV}$ and $m_{\PZ \PV}$ in the signal region are shown in Fig.~\ref{fig:signal2}. The data yields, together with the SM expectations for the different processes, are given in Table~\ref{tab:sel_yields15}. A nonzero aQGC enhances the production cross section at large masses of the $\PV\PV$ system with respect to the SM prediction, as can be seen in Fig.~\ref{fig:signal2}. The observed number of data events with $m_{\PV \PV}>1500\GeV$ is 3 (3) compared to the predicted SM background yield of $6.4\pm1.5$ ($2.6\pm1.3$) in the $\PW\PV$ ($\PZ\PV$) channel.

\begin{table*}[h!]
\centering
\topcaption{
Observed and expected lower and upper 95\% \CL limits on the parameters of the quartic operators S0, S1, M0, M1, M6, M7, T0, T1, and T2 in $\PW \PV$ and $\PZ \PV$ channels. The last two columns show the observed and expected limits for the combination of the $\PW \PV$ and $\PZ \PV$ channels.
}
\label{tab:VBS_aQGC}
\resizebox{\textwidth}{!}{
\begin{tabular}{ccccccc}
\hline
& Observed ($\PW \PV$)  & Expected ($\PW \PV$) & Observed ($\PZ \PV$) & Expected ($\PZ \PV$) & Observed & Expected  \\
& (TeV$^{-4}$)   & (TeV$^{-4}$)  & (TeV$^{-4}$)   & (TeV$^{-4}$) & (TeV$^{-4}$)   & (TeV$^{-4}$)  \\
\hline
$f_{\text{S0}} / \Lambda^4$  & $[ -2.7, 2.7]$ & $[ -4.2, 4.2]$ & $[ -40, 40]$ & $[ -31, 31]$  & $[ -2.7, 2.7]$ & $[ -4.2, 4.2]$ \\
$f_{\text{S1}} / \Lambda^4$  & $[-3.3, 3.4]$ & $[-5.2, 5.2]$ & $[-32, 32]$ & $[-24, 24]$  & $[-3.4, 3.4]$ & $[-5.2, 5.2]$ \\
$f_{\text{M0}} / \Lambda^4$  & $[-0.69, 0.69]$ & $[-1.0, 1.0]$ & $[-7.5, 7.5]$ & $[-5.3, 5.3]$  & $[-0.69, 0.70]$ & $[-1.0, 1.0]$ \\
$f_{\text{M1}} / \Lambda^4$  & $[ -2.0, 2.0]$ & $[ -3.0, 3.0]$ &  $[ -22, 23]$ & $[-16, 16]$  & $[ -2.0, 2.1]$ & $[ -3.0, 3.0]$ \\
$f_{\text{M6}} / \Lambda^4$  & $[-1.4, 1.4]$ & $[-2.0, 2.0]$  & $[-15, 15]$ & $[-11, 11]$ & $[-1.3, 1.3]$ & $[-1.4, 1.4]$ \\
$f_{\text{M7}} / \Lambda^4$  & $[-3.4, 3.4]$ & $[-5.1, 5.1]$ &  $[-35, 36]$ & $[-25, 26]$ & $[-3.4, 3.4]$ & $[-5.1, 5.1]$ \\
$f_{\text{T0}} / \Lambda^4$  & $[-0.12, 0.11]$ & $[-0.17, 0.16]$ & $[-1.4, 1.4]$ & $[-1.0, 1.0]$ & $[-0.12, 0.11]$ & $[-0.17, 0.16]$ \\
$f_{\text{T1}} / \Lambda^4$  & $[-0.12, 0.13]$ & $[-0.18, 0.18]$ &  $[-1.5, 1.5]$ & $[-1.0, 1.0]$ & $[-0.12, 0.13]$ & $[-0.18, 0.18]$  \\
$f_{\text{T2}} / \Lambda^4$  & $[-0.28, 0.28]$ & $[-0.41, 0.41]$ &  $[-3.4, 3.4]$ & $[-2.4, 2.4]$ &  $[-0.28, 0.28]$ & $[-0.41, 0.41]$ \\
\hline
\end{tabular}
}
\end{table*}

The observed and expected $95\%$ confidence level (\CL) lower and upper limits on the aQGC parameters  $f/\Lambda^4$, where $f$ is the dimensionless coefficient of the given operator and $\Lambda$ is the energy scale of new physics, are calculated using a modified frequentist approach with the \CLs criterion~\cite{Junk,Read} and asymptotic results for the test statistic~\cite{CLs}. The increase of the yield as a function of the aQGC exhibits a quadratic behavior, and a fitted parabolic function is used to interpolate between the discrete coupling parameters of the simulated signals. This is done for each bin of the mass distribution of the $\PW \PV$ or $\PZ \PV$ system. Table~\ref{tab:VBS_aQGC} shows the individual lower and upper limits obtained by setting all other aQGCs parameters to zero for the $\PW \PV$ and $\PZ \PV$ channels and their combination. These results give the most stringent constraints on the aQGC parameters for the S0, S1, M0, M1, M6, M7, T0, T1, and T2 operators. The effective field theory is not a complete model and the presence of nonzero aQGCs will violate tree-level unitarity at sufficiently high energy. It is important to note that the given limits do not include dipole form factors or other procedures to avoid unitarity violation~\cite{Eboli:2000ad}.

\begin{figure*}[!htbp]
	\centering
	\includegraphics[width=0.49\textwidth]{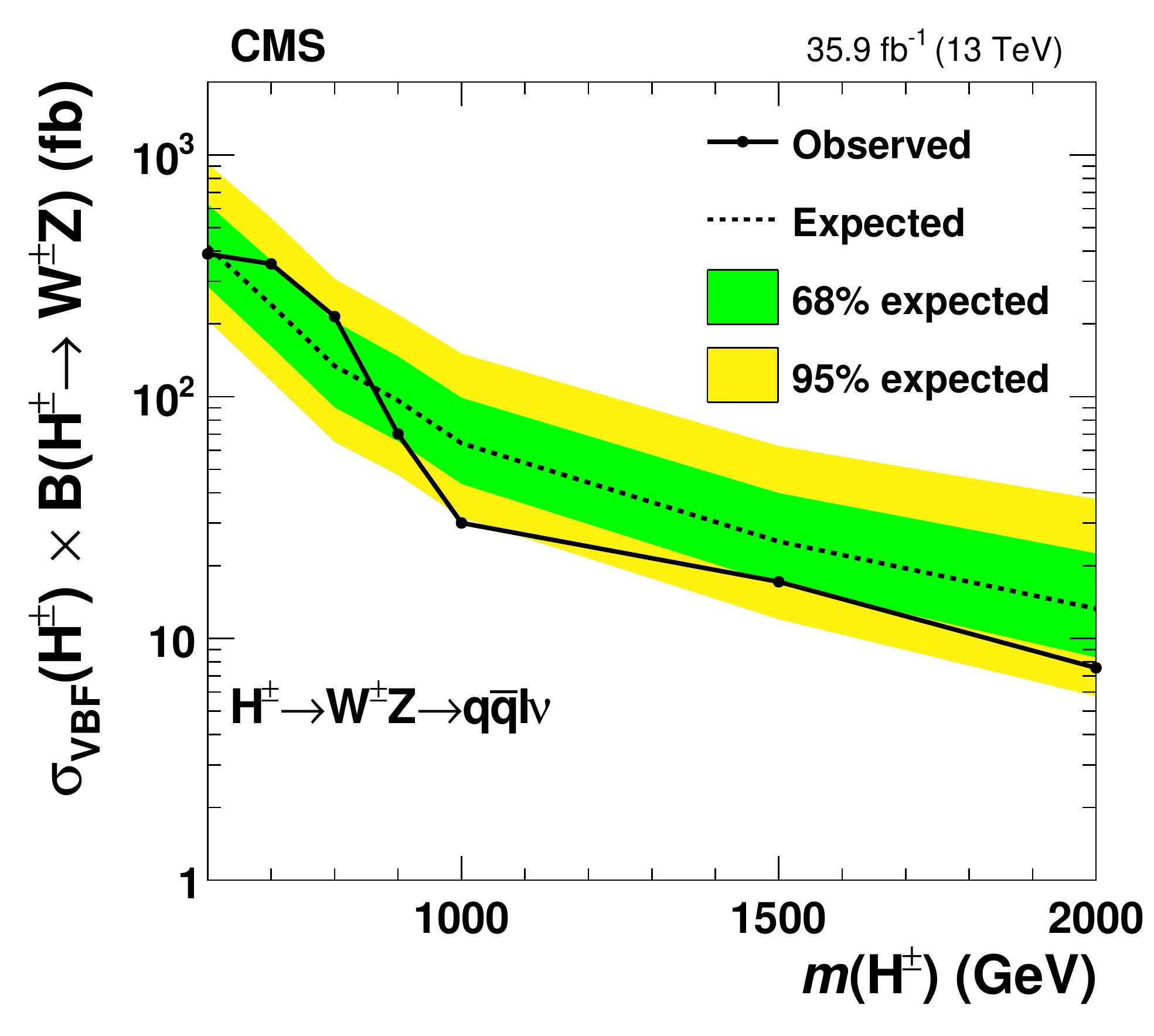}
	\includegraphics[width=0.49\textwidth]{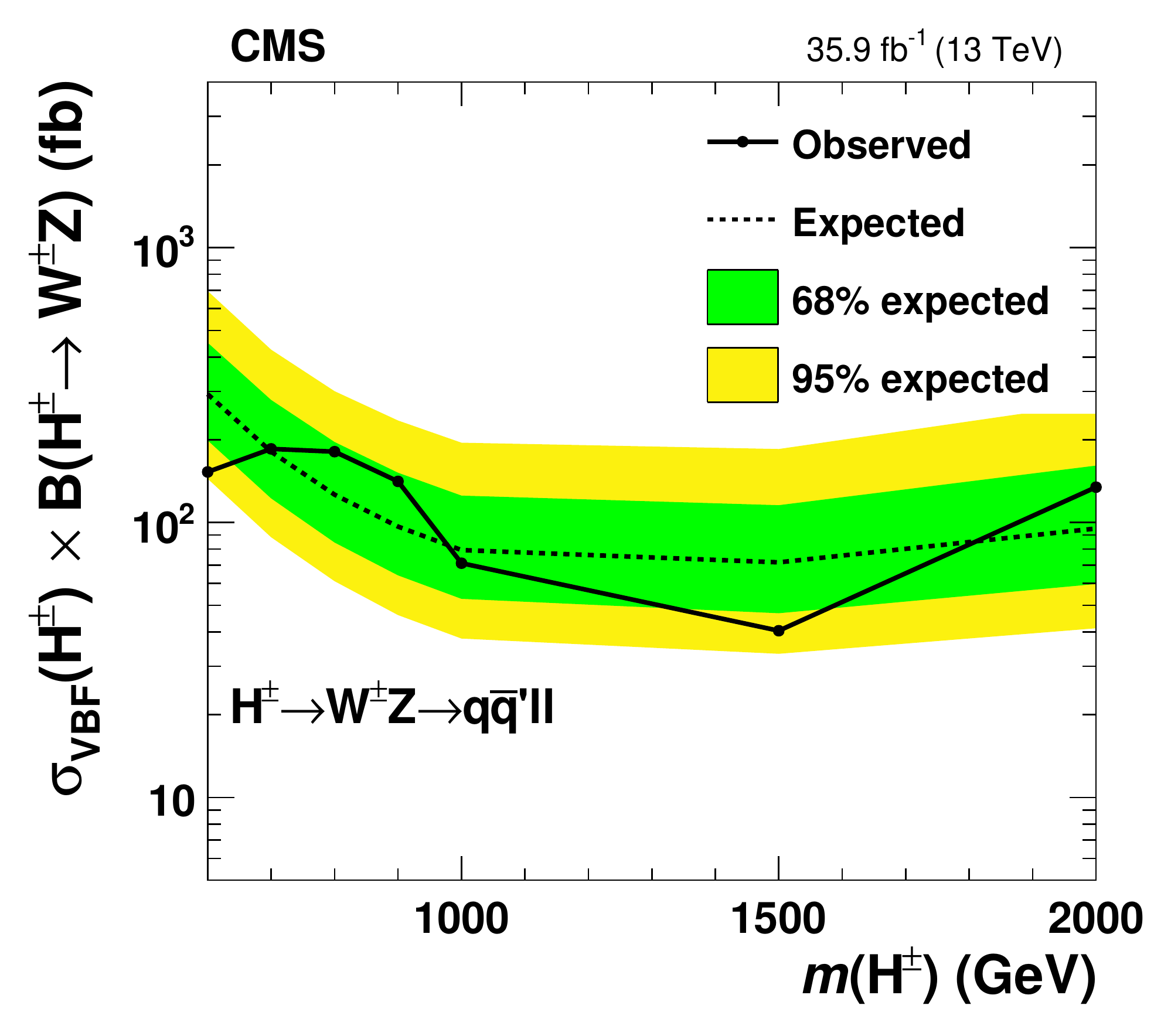}
	\includegraphics[width=0.49\textwidth]{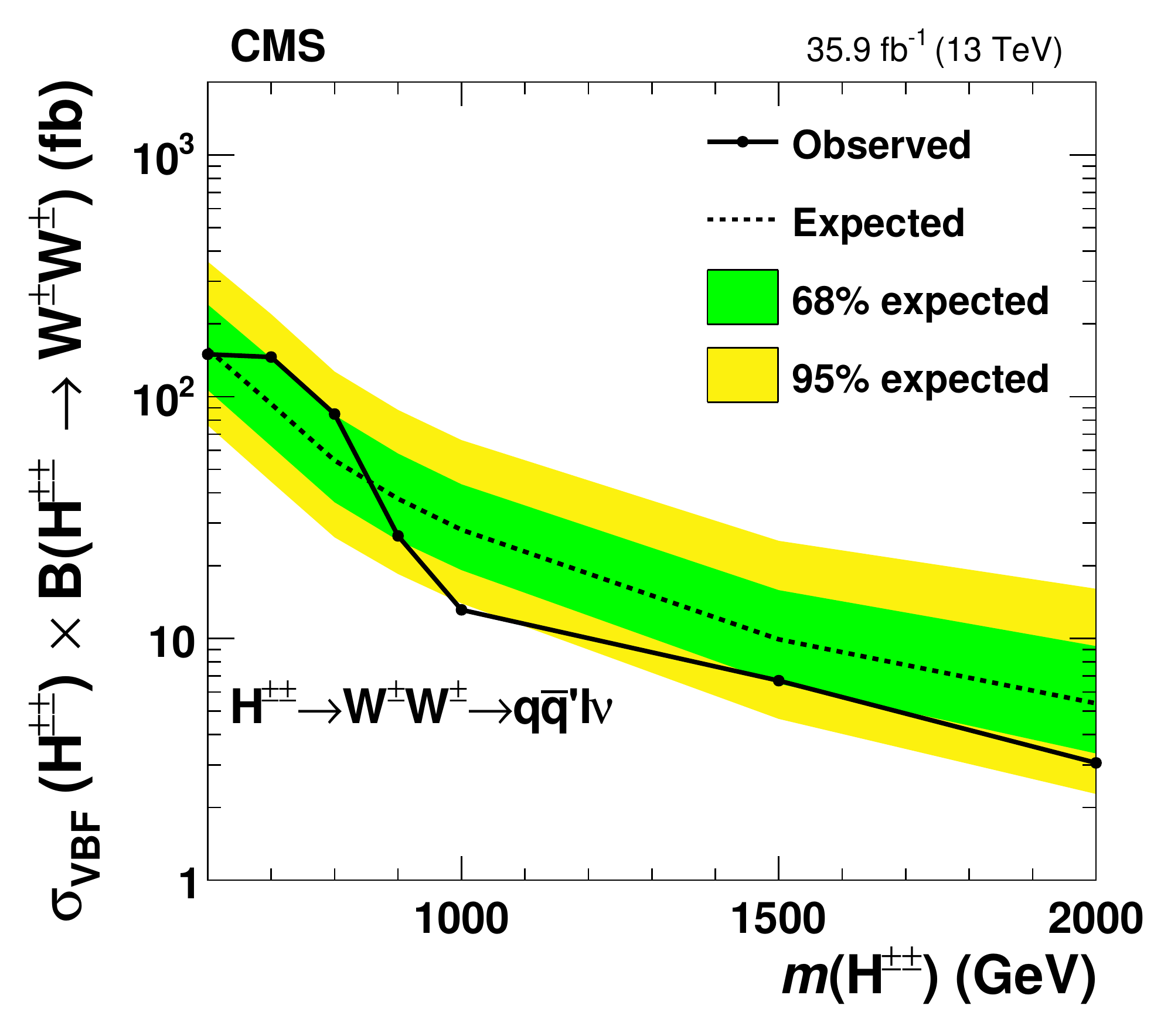}
	\includegraphics[width=0.49\textwidth]{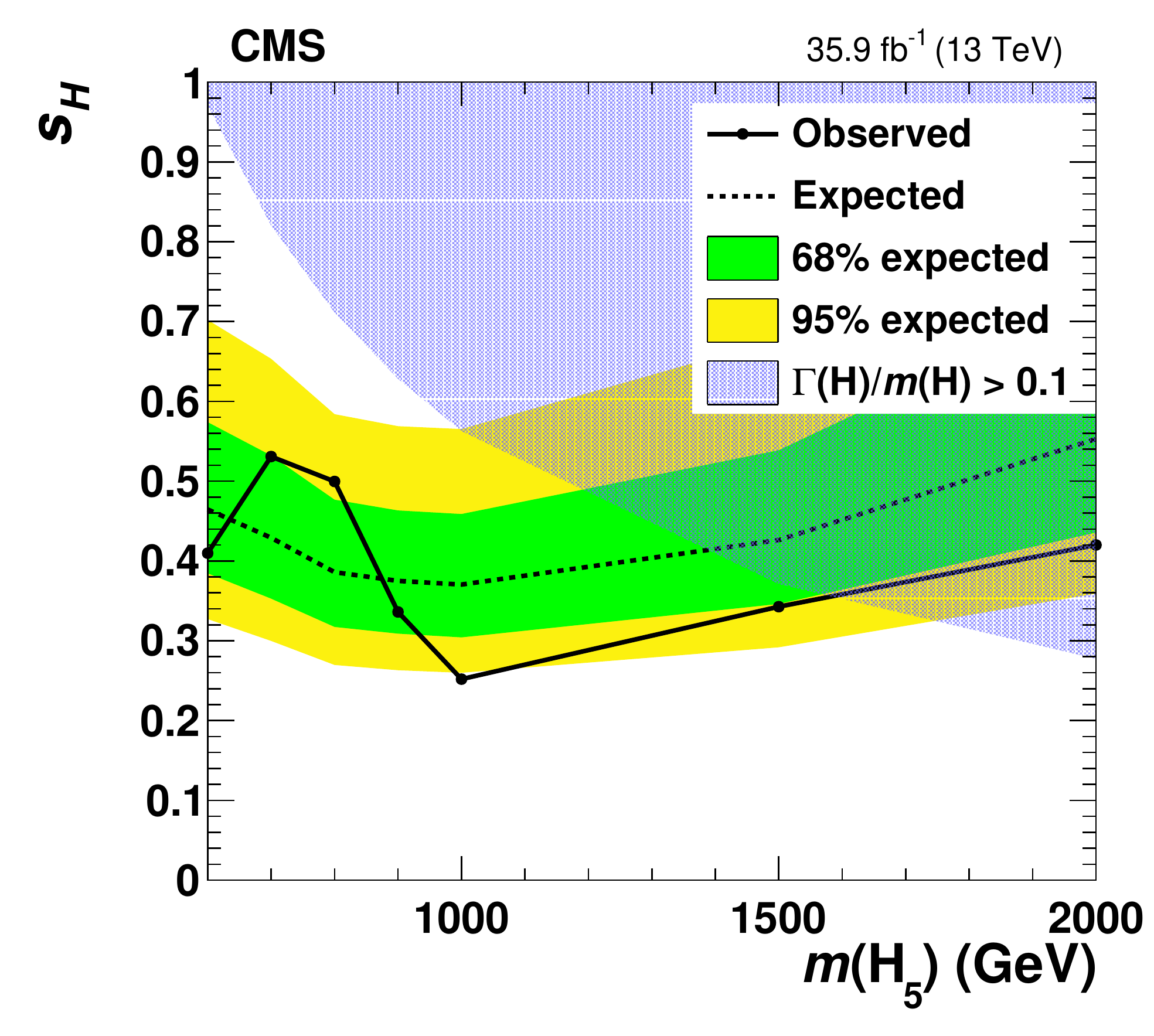}
	\caption{Expected and observed exclusion limits at the 95\% \CL as a function of $m(\PHpm)$ for $\sigma_\mathrm{VBF}(\PHpm) \, \mathcal{B}(\PHpm\to \PW^{\pm}\PZ)$ in the $\PW \PV$ (upper left) and $\PZ \PV$ (upper right) final states, for $\sigma_\mathrm{VBF}(\PHpmpm) \, \mathcal{B}(\PHpmpm\to \PW^{\pm}\PW^{\pm})$, as a function of $m(\PHpmpm)$ (lower left), and for $s_{\PH}$ in the GM model (lower right). The blue shaded area covers the theoretically disallowed parameter space~\cite{Zaro:2002500}.
	}
	\label{fig:limits}
\end{figure*}

Constraints on resonant charged Higgs boson production are also derived. The exclusion limits on the product of the charged Higgs boson cross section and branching fraction $\sigma_\mathrm{VBF}(\PHpm) \, \mathcal{B}(\PHpm\to \PW^{\pm}\PZ)$ at the 95\% \CL as a function of $m(\PHpm)$ for the $\PW^{\pm} \PV$ (upper left) and $\PZ \PV$ (upper right) channels, respectively, are shown in Fig.~\ref{fig:limits}. The exclusion limit on the doubly charged Higgs boson $\sigma_\mathrm{VBF}(\PHpmpm) \, \mathcal{B}(\PHpmpm\to \PW^{\pm}\PW^{\pm})$ at the 95\% \CL as a function of $m(\PHpmpm)$ for the $\PW \PV$ final state is also shown in the lower left panel in Fig.~\ref{fig:limits}. A small intrinsic width of 1\GeV is assumed for the $\PHpmpm$ and $\PHpm$ bosons. The combination of the model-independent exclusion limits constrains the $s_{\PH}$-$m(\PH_{5})$ plane by using the predicted cross sections at next-to-NLO accuracy in the GM model~\cite{Zaro:2002500}. The excluded $s_{\PH}$ values as a function of $m(\PH_{5})$ are shown in Fig.~\ref{fig:limits} (lower right).

\section{Summary}\label{sec:summary}
A search for anomalous electroweak production of $\PW \PW$, $\PW \PZ$, and $\PZ \PZ$ boson pairs in association with two jets in proton-proton collisions at the center-of-mass energy of $13\TeV$ was reported. The data sample corresponds to an integrated luminosity of $35.9\fbinv$ collected with the CMS detector at $13\TeV$. Final states with one or two leptons and a hadronically decaying $\PW/ \PZ$ boson, reconstructed as one large-radius jet, are considered. The contribution of the major background process $\PW$($\PZ$)+jets in the $\PW \PV$ ($\PZ \PV$) channel is evaluated with data control samples. No excess of events with respect to the SM background predictions is observed. Constraints on the quartic vector boson interactions in the framework of dimension-8 effective field theory operators are obtained. Stringent limits on the effective field theory operators S0, S1, M0, M1, M6, M7, T0, T1, and T2 are set. These are the first searches for anomalous electroweak production of $\PW \PW$, $\PW \PZ$, and $\PZ \PZ$ boson pairs in $\PW \PV$ and $\PZ \PV$ semi-leptonic channels at $13\TeV$. The limits improve the sensitivity of the current CMS fully leptonic results at $13\TeV$~\cite{Sirunyan:2017ret,Sirunyan:2017fvv,Sirunyan:2019ksz} by factors of up to seven, depending on the operator. The upper limits on VBF produced charged Higgs boson cross sections in the high-mass region extend the previous results at the LHC. The results are interpreted in the GM model where the observed limit excludes $s_{\PH}$ values greater than 0.53 for the $m(\PH_{5})$ range from 600 to 2000\GeV.

\begin{acknowledgments}
We congratulate our colleagues in the CERN accelerator departments for the excellent performance of the LHC and thank the technical and administrative staffs at CERN and at other CMS institutes for their contributions to the success of the CMS effort. In addition, we gratefully acknowledge the computing centers and personnel of the Worldwide LHC Computing Grid for delivering so effectively the computing infrastructure essential to our analyses. Finally, we acknowledge the enduring support for the construction and operation of the LHC and the CMS detector provided by the following funding agencies: BMBWF and FWF (Austria); FNRS and FWO (Belgium); CNPq, CAPES, FAPERJ, FAPERGS, and FAPESP (Brazil); MES (Bulgaria); CERN; CAS, MoST, and NSFC (China); COLCIENCIAS (Colombia); MSES and CSF (Croatia); RPF (Cyprus); SENESCYT (Ecuador); MoER, ERC IUT, PUT and ERDF (Estonia); Academy of Finland, MEC, and HIP (Finland); CEA and CNRS/IN2P3 (France); BMBF, DFG, and HGF (Germany); GSRT (Greece); NKFIA (Hungary); DAE and DST (India); IPM (Iran); SFI (Ireland); INFN (Italy); MSIP and NRF (Republic of Korea); MES (Latvia); LAS (Lithuania); MOE and UM (Malaysia); BUAP, CINVESTAV, CONACYT, LNS, SEP, and UASLP-FAI (Mexico); MOS (Montenegro); MBIE (New Zealand); PAEC (Pakistan); MSHE and NSC (Poland); FCT (Portugal); JINR (Dubna); MON, RosAtom, RAS, RFBR, and NRC KI (Russia); MESTD (Serbia); SEIDI, CPAN, PCTI, and FEDER (Spain); MOSTR (Sri Lanka); Swiss Funding Agencies (Switzerland); MST (Taipei); ThEPCenter, IPST, STAR, and NSTDA (Thailand); TUBITAK and TAEK (Turkey); NASU and SFFR (Ukraine); STFC (United Kingdom); DOE and NSF (USA).

\hyphenation{Rachada-pisek} Individuals have received support from the Marie-Curie program and the European Research Council and Horizon 2020 Grant, contract Nos.\ 675440 and 765710 (European Union); the Leventis Foundation; the A.P.\ Sloan Foundation; the Alexander von Humboldt Foundation; the Belgian Federal Science Policy Office; the Fonds pour la Formation \`a la Recherche dans l'Industrie et dans l'Agriculture (FRIA-Belgium); the Agentschap voor Innovatie door Wetenschap en Technologie (IWT-Belgium); the F.R.S.-FNRS and FWO (Belgium) under the ``Excellence of Science -- EOS" -- be.h project n.\ 30820817; the Beijing Municipal Science \& Technology Commission, No. Z181100004218003; the Ministry of Education, Youth and Sports (MEYS) of the Czech Republic; the Lend\"ulet (``Momentum") Program and the J\'anos Bolyai Research Scholarship of the Hungarian Academy of Sciences, the New National Excellence Program \'UNKP, the NKFIA research grants 123842, 123959, 124845, 124850, 125105, 128713, 128786, and 129058 (Hungary); the Council of Science and Industrial Research, India; the HOMING PLUS program of the Foundation for Polish Science, cofinanced from European Union, Regional Development Fund, the Mobility Plus program of the Ministry of Science and Higher Education, the National Science Center (Poland), contracts Harmonia 2014/14/M/ST2/00428, Opus 2014/13/B/ST2/02543, 2014/15/B/ST2/03998, and 2015/19/B/ST2/02861, Sonata-bis 2012/07/E/ST2/01406; the National Priorities Research Program by Qatar National Research Fund; the Programa Estatal de Fomento de la Investigaci{\'o}n Cient{\'i}fica y T{\'e}cnica de Excelencia Mar\'{\i}a de Maeztu, grant MDM-2015-0509 and the Programa Severo Ochoa del Principado de Asturias; the Thalis and Aristeia programs cofinanced by EU-ESF and the Greek NSRF; the Rachadapisek Sompot Fund for Postdoctoral Fellowship, Chulalongkorn University and the Chulalongkorn Academic into Its 2nd Century Project Advancement Project (Thailand); the Welch Foundation, contract C-1845; and the Weston Havens Foundation (USA).
\end{acknowledgments}

\bibliography{auto_generated}

\cleardoublepage \appendix\section{The CMS Collaboration \label{app:collab}}\begin{sloppypar}\hyphenpenalty=5000\widowpenalty=500\clubpenalty=5000\vskip\cmsinstskip
\textbf{Yerevan Physics Institute, Yerevan, Armenia}\\*[0pt]
A.M.~Sirunyan, A.~Tumasyan
\vskip\cmsinstskip
\textbf{Institut für Hochenergiephysik, Wien, Austria}\\*[0pt]
W.~Adam, F.~Ambrogi, E.~Asilar, T.~Bergauer, J.~Brandstetter, M.~Dragicevic, J.~Erö, A.~Escalante~Del~Valle, M.~Flechl, R.~Frühwirth\cmsAuthorMark{1}, V.M.~Ghete, J.~Hrubec, M.~Jeitler\cmsAuthorMark{1}, N.~Krammer, I.~Krätschmer, D.~Liko, T.~Madlener, I.~Mikulec, N.~Rad, H.~Rohringer, J.~Schieck\cmsAuthorMark{1}, R.~Schöfbeck, M.~Spanring, D.~Spitzbart, W.~Waltenberger, J.~Wittmann, C.-E.~Wulz\cmsAuthorMark{1}, M.~Zarucki
% ...
\vskip\cmsinstskip
% ...
1:  Also at Vienna University of Technology, Vienna, Austria\\
% ...
\end{sloppypar}
\end{document}